**Author for correspondence:**
Christian Reidys
e-mail: cmr3hk@virginia.edu

# The site linkage spectrum of data arrays

Christopher Barrett[1], Andrei Bura[2], Fenix Huang[3] and Christian Reidys[4]

[1,2,3,4]Biocomplexity Institute, University of Virginia, 994 Research Park Boulevard, Charlottesville, 22911, VA, USA.
[1]Department of Computer Science, University of Virginia, 351 McCormick Road, Charlottesville, 22904, VA, USA.
[4]Department of Mathematics, University of Virginia, 141 Cabell Drive, Charlottesville, 22904, VA, USA.

A new perspective is introduced regarding the analysis of Multiple Sequence Alignments (MSA), representing aligned data defined over a finite alphabet of symbols. The framework is designed to produce a block decomposition of an MSA, where each block is comprised of sequences exhibiting a certain site-coherence. The key component of this framework is an information theoretical potential defined on pairs of sites (links) within the MSA. This potential quantifies the expected drop in variation of information between the two constituent sites, where the expectation is taken with respect to all possible sub-alignments, obtained by removing a finite, fixed collection of rows.

It is proved that the potential is zero for linked sites representing columns, whose symbols are in bijective correspondence and it is strictly positive, otherwise. It is furthermore shown that the potential assumes its unique minimum for links at which each symbol pair appears with the same multiplicity.

Finally, an application is presented regarding anomaly detection in an MSA, composed of inverse fold solutions of a fixed tRNA secondary structure, where the anomalies are represented by inverse fold solutions of a different RNA structure.

# 1. Introduction

Aligned sequence data can be represented as an equivalence class of matrices with entries from a finite symbol alphabet, where two matrices are equivalent if they differ by the labeling of their rows. Such an equivalence class is referred to as a multiple sequence alignment (MSA) [1].

MSAs have been studied extensively in computational biology [2,3], in the bio-physics of molecular folding [4,5], in phylogenetic tree reconstruction [6–8], hidden Markov profile modeling and secondary or tertiary molecular structure and function prediction [9,10], as well as in primer design, and biological data validation [11,12]. In viral phylogeny MSAs are instrumental to the construction of sequence partitions corresponding to lineages [13–15] and the derivation of their phylogenetic trees. In the fields of protein and RNA folding, MSAs are employed to enhance the quality of single sequence, energy based folding algorithms via consensus structures [16]. In this paper we present first results towards a novel framework for the analysis of MSAs. This work is based on the notion that blocks of related sequences (rows) within an MSA, exhibit specific non-random linkages between their sites (columns). Within a single MSA multiple such blocks may co-exist and their site-linkage patterns can be quantified by information theoretic means. Molecular bonds in RNA sequences are a prime example of such linkages, representing the chemical coupling (typically between pairs) of nucleotides [17]. In general however, such linkages need not necessarily result from bio-physical interactions, and in fact could be truly information theoretical in nature. It is the variation of information associated with links, that can be leveraged in order to augment existing MSA analysis methods.

Our framework represents a version of sequence clustering, where blocks of sequences conforming to a fixed site-linkage pattern are constructed. However, the score by which these clusters are derived is a function of linkages and not based upon some metric or score on the sequences themselves. Specifically, we study the effect, that passing to subsets of the original sequence set, has on the variation of information measure between sites of the MSA. Our approach has the same objective as phylogeny: partitioning aligned data. The differentiating feature is on what basis the partitioning is achieved. Instead of considering local variations such as specific mutations observed at certain positions in the MSA, our framework considers the relational signature between the sites as a whole. Clustering of the original sequences can then be achieved by examining the information theoretic site-linkage spectrum, introduced in this paper.

In the context of molecular folding of proteins or RNA, the framework presented here differs in the following. Instead of deriving a single structure, or a partition function for the MSA [18], our method can be tailored to construct the most likely block-partition of the sequences in the MSA, where each block corresponds to a specific, most adequate structure, though it will not immediately reveal what this structure could be. Potentially, the methodology could augment the concept of molecular structure as employed in protein and RNA folding, by leveraging genuine information theoretic interactions exhibited among the sites of an MSA.

It may be informative to consider our approach from the perspective of Raman spectroscopy [19]. Raman spectroscopy derives inferences on vibrational modes of a molecule by inelastic photon scattering, the spectra of which provide insights about chemical structure, molecular interaction, phase, polymorphy or crystallinity. The MSA plays the role of the molecule with the sites assuming the roles of atoms. Information theoretic measurements are effected on the linkages between sites. The analogue of photon scattering are measurements quantifying the effect on the variation of information distance [20] between the sites as blocks of sequences are removed from the original MSA. This line of work leads to an information theoretic notion of potential. The collection of these measurements across links is organized into a site linkage spectrum of the MSA, and the main results in this paper establish first, basic mathematical properties. In particular, the ground state of this spectrum is unique, non-zero, and assumed exclusively by links between sites in which symbols appear with uniform probability. As in Raman spectroscopy, we are ultimately interested in determining the main "modes" of the MSA, which is to say, we aim to identify sequence blocks that are compatible with a specific relational structure.

The paper is organized as follows: in section 2 we introduce basic concepts and definitions. Section 3 introduces the information theoretic potential and establishes basic properties. In section 4 we show that the ground state of the potential is uniquely assumed for links representing two columns in which any pair of symbols appears with the same frequency. Finally, section 5 introduces the site-linkage spectrum of an MSA and presents an example of how this concept applies in the context of RNA folding data.

## 2. Some basic facts

In the following we shall fix an ordering of the sequences (rows) of an MSA, which allows to reference the removal of single rows or blocks of rows, via indices. An MSA consisting of $k$ sequences, each of length $n$, over an alphabet of symbols $\mathcal{A}$, together with a fixed row ordering, is called *a system* of size $k$ and length $n$

$$\Sigma := \{\sigma^h = (y_{h,1}, \ldots, y_{h,n}) \mid y_{h,i} \in \mathcal{A}, 1 \leq h \leq k\}.$$

We refer to a column index $i \in \{1, \ldots, n\}$ as a *site*. We can associate to each site of $\Sigma$ a random variable $Y_i$, with states in $\mathcal{A}$, and with distribution $p_{y_i}$ given by the relative frequency with which the symbol $y_i$ appears in the column of $\Sigma$ indexed by $i$. Similarly, a pair of sites $(i, j)$ in $\Sigma$, are called a *link*, and have associated to them a random variable $Y_{i,j} = (Y_i, Y_j)$ with states in $\mathcal{A}^2$, and with distribution $p_{y_i, y_j}$ given by the relative frequency with which the pair of symbols $(y_i, y_j)$ appears in the column pair of $\Sigma$ indexed by $(i, j)$. For convenience we will denote columns or pairs of columns by the same labels as their associated random variables. Finally, for any row-index set $M \subseteq \{1, \cdots, k\}$, we can define the $\Sigma$-subsystem

$$\Sigma^M := \{\sigma^h \in \Sigma \mid h \notin M\} \subseteq \Sigma,$$

obtained from $\Sigma$ by deleting all rows with indexes in $M$, where by construction we have $\Sigma = \Sigma^\varnothing$.

Consider the case of a column pair $Y_{i,j}$, i.e., the case of a single link. Let $Z \in \{Y_i, Y_j, Y_{i,j}\}$ and fix $1 \leq s \leq k$. Let $\mathcal{P}_s := \{M \subseteq \{1, \ldots, k\} \mid |M| = s\}$ be a discrete sample space of outcomes, where the probability of selecting the subset $M$ is $p_M := \binom{k}{s}^{-1}$. With respect to this we define the random variable $\mathbb{Z}$ whose states are

$$\mathbb{Z}(M) := Z^M.$$

Suppose, $z$ denotes respectively, a fixed symbol or a fixed pair of symbols, i.e., $z \in \{y_i, y_j, (y_i, y_j)\}$. If $z$ is found in a row of $Y_{i,j}$ indexed by elements of $M$, then we write $z \sqsubset M$, and we write $z \not\sqsubset M$ otherwise. Furthermore, we denote the absolute frequency of $z$ in $M$ by

$$M(z) := |\{x \in M \mid z \sqsubset \{x\}\}|,$$

and the absolute frequency of $z$ in $Z$ by $\nu_z$. We note that for a fixed $M \in \mathcal{P}_s$, the object $\mathbb{Z}(M)$ is on the one hand, a state of the random variable $\mathbb{Z}$, but on the other hand, $\mathbb{Z}(M)$ is also a random variable in its own right. In view of

$$\sum_{z \sqsubset M} \frac{\nu_z - M(z)}{k-s} + \sum_{z \not\sqsubset M} \frac{\nu_z}{k-s} = \sum_{z \sqsubset M} \frac{\nu_z}{k-s} + \sum_{z \not\sqsubset M} \frac{\nu_z}{k-s} - \sum_{z \sqsubset M} \frac{M(z)}{k-s} = \frac{k}{k-s} - \frac{s}{k-s} = 1,$$

the random variable $\mathbb{Z}(M)$, for fixed $M$, has states $z$ and distribution

$$p_z^M = \begin{cases} p_z^- := \frac{\nu_z - M(z)}{k-s} & \text{for } z \sqsubset M \\ p_z^+ := \frac{\nu_z}{k-s} & \text{for } z \not\sqsubset M. \end{cases} \tag{2.1}$$

In particular, $\mathbb{Y}_i(M) = Y_i^M$, $\mathbb{Y}_j(M) = Y_j^M$ and $\mathbb{Y}_{i,j}(M) = Y_{i,j}^M$. The original system $Y_{i,j}$ as well as any sub-system $Y_{i,j}^M = (Y_i^M, Y_j^M)$, features a variation of information distance [21]

$$\mathfrak{D}(Y_{i,j}^M) := 2H(Y_{i,j}^M) - H(Y_i^M) - H(Y_j^M), \tag{2.2}$$

where $H(X)$ denotes the Shannon Entropy [22] of the discrete random variable $X$ and $H(X|Y)$ will denote the conditional entropy of $X$, given the random variable $Y$. Intuitively, the $\mathfrak{D}$-distance of a link quantifies (symmetrically) the amount of information that can be inferred from one site by knowing the other. We set

$$\mathfrak{D} \circ \mathbb{Y}_{i,j} : \mathcal{P}_s \to \mathbb{R}, \quad \mathfrak{D}(\mathbb{Y}_{i,j}(M)) := \mathfrak{D}(Y_{i,j}^M).$$

The expectation value of this random variable measures the average $\mathfrak{D}$-distance of sub-systems of $Y_{i,j}$ of size $k - s$, when these systems were each obtained by uniformly deleting $s$ rows from $Y_{i,j}$,

$$\mathbb{E}_s[\mathfrak{D}(\mathbb{Y}_{i,j})] := \frac{1}{\binom{k}{s}} \sum_{M \in \mathcal{P}_s} \mathfrak{D}(\mathbb{Y}_{i,j}(M)). \tag{2.3}$$

In particular, $\mathfrak{D}(Y_{i,j}) = \sum_{\varnothing} \mathfrak{D}(\mathbb{Y}_{i,j}(\varnothing)) = \mathbb{E}_0[\mathfrak{D}(\mathbb{Y}_{i,j})]$. We next consider the drop in $\mathfrak{D}$-distance across $s$-block row removals at the link $(i, j)$:

$$\mathcal{Y}_{i,j} : \mathcal{P}_s \to \mathbb{R}, \quad \mathcal{Y}_{i,j}(M) := \mathfrak{D}(Y_{i,j}) - \mathfrak{D}(\mathbb{Y}_{i,j}(M)). \tag{2.4}$$

We call its expectation $\mathbb{E}_s[\mathcal{Y}_{i,j}] = \mathfrak{D}(\mathbb{Y}_{i,j}) - \mathbb{E}_s[\mathfrak{D}(\mathbb{Y}_{i,j})]$, the *link-potential* at $(i, j)$. In view of eq.( 2.3) this can also be written as $\mathbb{E}_s[\mathcal{Y}_{i,j}] = \mathbb{E}_0[\mathfrak{D}(\mathbb{Y}_{i,j})] - \mathbb{E}_s[\mathfrak{D}(\mathbb{Y}_{i,j})]$.

For $s = 1$, we set $M = \{x\} \in \mathcal{P}_1$ and shall write $Y_{i,j}^x = \mathbb{Y}_{i,j}(\{x\})$ for $x \in \{1, \ldots, k\}$. In this case, letting $X$ be a random variable with states $x \in \{1, \ldots, k\}$ and uniform distribution $p_x = k^{-1}$, we have the following interpretation of the potential in terms of Shannon Mutual Information [23]:

**Proposition 2.1.**

$$\mathbb{E}_1[\mathcal{Y}_{i,j}] = 2I(Y_{i,j}, X) - I(Y_i, X) - I(Y_j, X). \tag{2.5}$$

*Proof.* In view of eq. (2.1), we can define $S$ as a random variable with states being pairs $(z, x)$ and distribution $p_{(z,x)} = p_x p_z^x$. We have

$$\sum_z \sum_x \frac{1}{k} p_z^x = \sum_z \frac{1}{k} \left[ \nu_z p_z^- + (k - \nu_z) p_z^+ \right] = \sum_z p_z p_z^- + (1 - p_z) p_z^+ = \sum_z p_z = 1,$$

whence $p_{(z,x)}$ is indeed a probability distribution. The $p_{(z,x)}$-marginals are given by

$$\sum_x p_{(z,x)} = \sum_x P(S = (z, x)) \quad = \quad \sum_x \frac{1}{k} p_z^x = p_z p_z^- + (1 - p_z) p_z^+ = p_z$$

$$\sum_z p_{(z,x)} = \sum_z P(S = (z, x)) \quad = \quad \frac{1}{k} \sum_z p_z^x = \frac{1}{k} \left[ \sum_{z \sqsubset x} \frac{\nu_z - 1}{k - 1} + \sum_{z \not\sqsubset x} \frac{\nu_z}{k - 1} \right] = \frac{1}{k} = p_x.$$

Hence $p_{(z,x)}$ represents a joint distribution with marginals $p_z$ and $p_x$, respectively. Accordingly,

$$H(Z \mid X) = -\sum_z \sum_x p_{(z,x)} \ln \left( \frac{p_{(z,x)}}{p_x} \right) = -\frac{1}{k} \sum_z \sum_x p_z^x \ln \left( p_z^x \right) = -\frac{1}{k} \sum_x H(\mathbb{Z}(x)).$$

In view of eq. (2.2)

$$\begin{aligned}
\mathbb{E}_1[\mathfrak{D}(\mathbb{Y}_{i,j})] \quad &= \quad \frac{1}{k} \sum_x \mathfrak{D}(Y_{i,j}^x) \\
&= \quad 2\frac{1}{k} \sum_x H(\mathbb{Y}_{i,j}(x)) - \frac{1}{k} \sum_x H(\mathbb{Y}_i(x)) - \frac{1}{k} \sum_x H(\mathbb{Y}_j(x)) \\
&= \quad 2H(Y_{i,j} \mid X) - H(Y_i \mid X) - H(Y_j \mid X)
\end{aligned}$$

and as a result,

$$\mathbb{E}_1[\mathcal{Y}_{i,j}] = 2[H(Y_{i,j}) - H(Y_{i,j} \mid X)] - [H(Y_i) - H(Y_i \mid X)] - [H(Y_j) - H(Y_j \mid X)].$$

Since $I(Y, X) = H(Y) - H(Y|X)$, eq. (2.5) follows. $\qquad\square$

## 3. The link-potential

Let $\beta\colon \mathcal{A} \to \mathcal{A}$ be a bijection of the symbol set and let $Y_{i,j}$ be a column pair. In case there exists some bijection $\beta$ such that $y_{h,j} = \beta(y_{h,i})$, which we denote by abuse of notation as $Y_j = \beta(Y_i)$, we call the link $(i, j)$ *pure*.

**Theorem 3.1.** *Let $Y_{i,j}$ be a pair of columns of size $k$. Then*

$$\mathbb{E}_s\left[\mathcal{Y}_{i,j}\right] = \begin{cases} 0 & \text{for pure links } (i, j), \\ r \in \mathbb{R}^+ & \text{otherwise.} \end{cases}$$

*Furthermore, for links that are not pure, there always exists a subset of rows, whose removal strictly decreases the $\mathfrak{D}$-distance.*

The proof of Theorem 3.1 will be based on Lemma 3.1 and obtained by induction on $s = |M|$.

**Lemma 3.1.** *Let $Y_{i,j}$ be a pair of columns of size $k$ whose associated link, $(i, j)$ is not pure. Let $G(z) = z(z-1)\ln\left[1 + \frac{1}{z-1}\right]$ and*

$$\lambda(Y_{i,j}) = \frac{1}{k}\left[\sum_{y_i} G(\nu_{y_i}) + \sum_{y_j} G(\nu_{y_j}) - 2\sum_{(y_i, y_j)} G(\nu_{y_i, y_j})\right].$$

*Then*

$$\mathbb{E}_1[\mathcal{Y}_{i,j}] = \frac{\lambda(Y_{i,j})}{(k-1)} > 0.$$

*Furthermore there always exists a row whose removal strictly decreases the $\mathfrak{D}$-distance.*

*Proof.* Let $z \in \{y_i, y_j, (y_i, y_j)\}$ and $\mathbb{Z}(x) = Z^x$. Then $\mathbb{Z}(x)$ has distribution $\Pr(\mathbb{Z}(x) = z) := p_z^x$, for

$$p_z^x = \begin{cases} p_z^+ := \frac{\nu_z}{k-1} & \text{for } z \not\sqsubset x \\ p_z^- := \frac{\nu_z - 1}{k-1} & \text{for } z \sqsubset x \end{cases}$$

Furthermore, let

$$F(Z) = \sum_z p_z p_z^- \ln\left[\frac{p_z^-}{p_z}\right] + \sum_z (1 - p_z) p_z^+ \ln\left[\frac{p_z^+}{p_z}\right].$$

*Claim* 1.

$$\mathbb{E}_1[\mathcal{Y}_{i,j}] = 2F(Y_{i,j}) - [F(Y_i) + F(Y_j)].$$

To prove Claim 1, note that by eq. (2.2),

$$\mathfrak{D}(Y_{i,j}) = 2H(Y_{i,j}) - H(Y_i) - H(Y_j)$$

and eq. (2.3) implies

$$\mathbb{E}_1[\mathfrak{D}(\mathbb{Y}_{i,j})] = \frac{1}{k}\left[\sum_x 2H(\mathbb{Y}_{i,j}(x)) - H(\mathbb{Y}_i(x)) - H(\mathbb{Y}_j(x))\right].$$

Then $\mathfrak{D}(Y_{i,j}) - \mathbb{E}_1[\mathfrak{D}(\mathbb{Y}_{i,j})]$ can be written as a sum of terms of the form $[H(Z) - \frac{1}{k}\sum_x H(Z^x)]$ for $Z = Y_i, Y_j, Y_{i,j}$, and we arrive at

$$\mathbb{E}_1[\mathcal{Y}_{i,j}] = 2[H(Y_{i,j}) - \frac{1}{k}\sum_x H(Y_{i,j}^x)] - [H(Y_i) - \frac{1}{k}\sum_x H(Y_i^x)] - [H(Y_j) - \frac{1}{k}\sum_x H(Y_j^x)].$$

In view of

$$p_z p_z^- + (1 - p_z)p_z^+ = \frac{\nu_z}{k}\frac{\nu_z - 1}{k - 1} + \left(1 - \frac{\nu_z}{k}\right)\frac{\nu_z}{k - 1} = p_z, \qquad (3.1)$$

we derive

$$
\begin{aligned}
H(Z) - \frac{1}{k}\sum_x H(Z^x) &= -\left[\sum_z p_z \ln(p_z) - \frac{1}{k}\sum_x \sum_z p_z^x \ln(p_z^x)\right]\\
&= -\left[\sum_z [p_z p_z^- + (1-p_z)p_z^+]\ln(p_z) - \frac{1}{k}\left[\sum_{\substack{(z,x)\\ z\sqsubset x}} p_z^- \ln(p_z^-) + \sum_{\substack{(z,x)\\ z\not\sqsubset x}} p_z^+ \ln(p_z^+)\right]\right].
\end{aligned}
$$

Fixing $z$ and summing over all row-removals $x$, we encounter two cases: either $z \sqsubset x$ and then accounts for a term $p_z^-$, in which case over all rows this occurs exactly $\nu_z$ times, or $z \not\sqsubset x$, in which case a term $p_z^+$ is produced with multiplicity $(k-\nu_z)$. Accordingly,

$$
\begin{aligned}
\frac{1}{k}\left[\sum_{z\sqsubset x} p_z^- \ln(p_z^-) + \sum_{z\not\sqsubset x} p_z^+ \ln(p_z^+)\right] &= \frac{1}{k}\left[\sum_z \nu_z p_z^- \ln(p_z^-) + \sum_z (k-\nu_z)p_z^+ \ln(p_z^+)\right]\\
&= \left[\sum_z p_z p_z^- \ln(p_z^-) + \sum_z (1-p_z)p_z^+ \ln(p_z^+)\right].
\end{aligned}
$$

The claim then follows from

$$
H(Z) - \frac{1}{k}\sum_x H(Z^x) = \sum_z p_z p_z^- \ln\left[\frac{p_z^-}{p_z}\right] + \sum_z (1-p_z)p_z^+ \ln\left[\frac{p_z^+}{p_z}\right] = F(Z).
$$

*Claim* 2.

$$
\mathbb{E}_1[\mathcal{Y}_{i,j}] = \frac{\lambda(Y_{i,j})}{(k-1)} > 0.
$$

To prove Claim 2, we compute

$$
\begin{aligned}
\sum_z p_z^+(1-p_z)\ln\frac{p_z^+}{p_z} &= \sum_z (1-p_z)\left(\frac{\nu_z}{k-1}\right)\ln\left[\frac{k}{k-1}\right]\\
\sum_z p_z^- p_z \ln\frac{p_z^-}{p_z} &= \sum_z p_z^- p_z \left[\ln\left[\frac{\nu_z-1}{\nu_z}\right] + \ln\left[\frac{k}{k-1}\right]\right].
\end{aligned}
$$

By eq. (3.1) we can conclude

$$
\sum_z p_z^+(1-p_z) + \sum_z p_z^- p_z = \sum_z p_z = 1,
$$

whence

$$
\sum_z p_z^+(1-p_z)\ln\left[\frac{k}{k-1}\right] + \sum_z p_z^- p_z \ln\left[\frac{k}{k-1}\right] = \ln\left[\frac{k}{k-1}\right].
$$

Consequently

$$
F(Z) = \ln\left[\frac{k}{k-1}\right] + \sum_z p_z^- p_z \ln\left[\frac{\nu_z-1}{\nu_z}\right]
$$

and since $\ln\left[\frac{\nu_z-1}{\nu_z}\right] = -\ln\left[\frac{\nu_z}{\nu_z-1}\right]$, we can write

$$
2F(Y_{i,j}) - [F(Y_i) + F(Y_j)] =
$$

$$
= \sum_{y_i} p_{y_i}^- \cdot p_{y_i} \ln\left[\frac{\nu_{y_i}}{\nu_{y_i}-1}\right] + \sum_{y_j} p_{y_j}^- \cdot p_{y_j} \ln\left[\frac{\nu_{y_j}}{\nu_{y_j}-1}\right] - 2\sum_{(y_i,y_j)} p_{y_i,y_j}^- \cdot p_{y_i,y_j} \ln\left[\frac{\nu_{y_i,y_j}}{\nu_{y_i,y_j}-1}\right].
$$

Setting

$$
\lambda(Y_{i,j}) = \frac{1}{k}\left[\sum_{y_i} G(\nu_{y_i}) + \sum_{y_j} G(\nu_{y_j}) - 2\sum_{(y_i,y_j)} G(\nu_{y_i,y_j})\right], \tag{3.2}
$$

we arrive at

$$2F(Y_{i,j}) - [F(Y_i) + F(Y_j)] = \frac{\lambda(Y_{i,j})}{k-1}. \tag{3.3}$$

We proceed by analyzing the RHS of eq. (3.3). Its terms represent summations over the symbols $y_i, y_j$ and symbol pairs $(y_i, y_j)$. Fixing a row $x$, determines a unique symbol pair $(y_i, y_j)$ as well as its marginals. Let

$$g(z) = (z-1)\ln\left[1 + \frac{1}{z-1}\right]. \tag{3.4}$$

Employing $g$, the expression for $\lambda(Y_{i,j})$ becomes

$$\lambda(Y_{i,j}) = \frac{1}{k}\left[\sum_{\substack{(y_i,x) \\ y_i \sqsubset x}} g(\nu_{y_i}) + \sum_{\substack{(y_j,x) \\ y_j \sqsubset x}} g(\nu_{y_j}) - 2\sum_{\substack{((y_i,y_j),x) \\ (y_i,y_j)\sqsubset x}} g(\nu_{y_i,y_j})\right]. \tag{3.5}$$

In eq. (3.5), we encounter for fixed $x$, terms: $g(\nu_{y_i})$, $g(\nu_{y_j})$ and $g(\nu_{y_i,y_j})$, where $(y_i, y_j) \sqsubset x$. To conclude our claim, it thus suffices to show

$$\forall (y_i, y_j) \sqsubset x; \quad g(\nu_{y_i}) + g(\nu_{y_j}) - 2g(\nu_{y_i,y_j}) > 0.$$

We compute

$$g'(z) = \ln\left[1 + \frac{1}{z-1}\right] + \frac{1}{1 + \frac{1}{z-1}} \cdot \frac{-1}{(z-1)}$$

and setting $\epsilon = \frac{1}{z-1} \leq 1$ we obtain, in view of $\ln(1+\epsilon) \geq \epsilon - \frac{\epsilon^2}{2}$,

$$\left(\epsilon - \frac{\epsilon^2}{2}\right) - \frac{\epsilon}{1+\epsilon} = \epsilon\left(1 - \frac{\epsilon}{2} - \frac{1}{1+\epsilon}\right).$$

Since $\epsilon \geq 0$, it remains to verify that $\left(1 - \frac{\epsilon}{2} - \frac{1}{1+\epsilon}\right)$ is strictly positive. This is equivalent to

$$\forall\, 0 < \epsilon \leq 1; \qquad (1+\epsilon)\left(1 - \frac{1}{2}\epsilon\right) = 1 + \epsilon - \frac{1}{2}\epsilon - \frac{1}{2}\epsilon^2 > 1$$

and as a result for $z \in (1, \infty)$, $g'(z) > 0$ holds. That is, $g(z)$ is monotonously increasing for $z > 1$. By assumption $(i, j)$ is not pure, whence there exists no bijection $\beta \colon \mathcal{A} \to \mathcal{A}$ such that $Y_j = \beta(Y_i)$. Accordingly, some $p_{y_i}, p_{y_j}$ are strictly larger than the joint probability $p_{y_i, y_j}$. Consequently, some $\nu_{y_i}$ and $\nu_{y_j}$ are strictly larger than $\nu_{(y_i, y_j)}$, whence $g(\nu_{y_i}) + g(\nu_{y_j}) - 2g(\nu_{y_i, y_j}) > 0$. In case $(y_i, y_j)$ appears only once we also obtain positivity, since

$$g(1) = \lim_{z \to 1}(z-1)\ln\left[1 + \frac{1}{z-1}\right] = 0.$$

As a result we have established $\lambda(Y_{i,j}) > 0$, which proves the claim.

Since the expectation is strictly positive, there always exists some row, $x$, such that

$$\mathfrak{D}(Y_{i,j}) - \mathfrak{D}(Y_{i,j}^x) \geq \mathbb{E}_1[\mathfrak{D}(Y_{i,j}) - \mathfrak{D}(\mathbb{Y}_{i,j})] = \mathbb{E}_1[\mathcal{Y}_{i,j}] = \frac{\lambda(Y_{i,j})}{(k-1)} > 0.$$

and the second assertion follows, completing the proof of the lemma. $\qquad \square$

We are now in position to prove Theorem 3.1.

*Proof.* Firstly, if the link $(i, j)$ is pure, then there exists a bijection $\beta$ s.t. $Y_j = \beta(Y_i)$ and $\mathfrak{D}(Y_{i,j}) = \mathfrak{D}(Y_{i,j}^M) = 0$ for any $M \subset \{1, \ldots, k\}$, whence $\mathbb{E}_s[\mathcal{Y}_{i,j}] = 0$ as claimed.

It therefore suffices to consider the case of a link that is not pure. For any $1 \leq s \leq k$ we have in view of eq. (2.3):

$$\mathbb{E}_s[\mathfrak{D}(\mathbb{Y}_{i,j})] = \frac{1}{\binom{k}{s}} \sum_M \left[ 2H(\mathbb{Y}_{i,j}(M)) - H(\mathbb{Y}_i(M)) - H(\mathbb{Y}_j(M)) \right],$$

where

$$H(\mathbb{Z}(M)) = -\sum_z p_z^M \ln(p_z^M) \quad \text{and} \quad p_z^M = \begin{cases} \frac{\nu_z}{k-s} & \text{for } z \not\sqsubset M \\ \frac{\nu_z - M(z)}{k-s} & \text{for } z \sqsubset M. \end{cases}$$

We use induction on $s$ to show

$$\mathbb{E}_s[\mathcal{Y}_{i,j}] = \mathfrak{D}(Y_{i,j}) - \mathbb{E}_s[\mathfrak{D}(\mathbb{Y}_{i,j})] > 0.$$

Firstly, Lemma 3.1 establishes the induction basis:

$$\mathbb{E}_1[\mathcal{Y}_{i,j}] = \mathfrak{D}(Y_{i,j}) - \mathbb{E}_1[\mathfrak{D}(\mathbb{Y}_{i,j})] > 0.$$

Secondly, by induction hypothesis, we can assume

$$\mathbb{E}_{s-1}[\mathcal{Y}_{i,j}] = \mathfrak{D}(i,j) - \mathbb{E}_{s-1}[\mathfrak{D}(\mathbb{Y}_{i,j})] > 0. \tag{3.6}$$

We shall remove rows in two rounds: first a subset of rows, $N_0$ with $|N_0| = s - 1$, followed by the removal of a single row. Beginning with $N_0$, Lemma 3.1 implies for the subsequent removal of a single row

$$\forall N_0 \subset \{1, \ldots, k\}; \quad \mathfrak{D}(\mathbb{Y}_{i,j}(N_0)) - \mathbb{E}_1[\mathfrak{D}(\mathbb{Y}_{i,j}(N_0))] \geq 0, \tag{3.7}$$

where

$$\mathbb{E}_1[\mathfrak{D}(\mathbb{Y}_{i,j}(N_0))] = \frac{1}{k - (s-1)} \sum_x \mathfrak{D}(\mathbb{Y}_{i,j}(N_0)^x).$$

For $\mathbb{Y}_{i,j}(N_0)$, either the link $(i,j)$ is pure or Lemma 3.1 applies. In either case we have

$$\mathfrak{D}(\mathbb{Y}_{i,j}(N_0)) - \mathbb{E}_1[\mathfrak{D}(\mathbb{Y}_{i,j}(N_0))] \geq 0.$$

Eq. (3.7) holds for any fixed $N_0$ and taking the normalized sum over all $(s-1)$-subsets, $N \subset \{1, \ldots, k\}$ we arrive at

$$\frac{1}{\binom{k}{s-1}} \sum_N \mathfrak{D}(\mathbb{Y}_{i,j}(N)) - \left[ \frac{1}{\binom{k}{s-1}} \sum_N \mathbb{E}_1 \left[ (\mathfrak{D}(\mathbb{Y}_{i,j}(N)) \right] \right] \geq 0. \tag{3.8}$$

The first term of eq. (3.8) has the following interpretation

$$\frac{1}{\binom{k}{s-1}} \sum_N \mathfrak{D}(\mathbb{Y}_{i,j}(N)) = \mathbb{E}_{s-1} \left[ \mathfrak{D}(\mathbb{Y}_{i,j}) \right]$$

and for the second term of eq. (3.8) we observe

$$\frac{1}{\binom{k}{s-1}} \sum_N \left[ \frac{1}{k - (s-1)} \sum_x \mathfrak{D}(\mathbb{Y}_{i,j}(N)^x) \right] = \frac{(s-1)!(k+1-s)!}{k!(k+1-s)} \sum_{(N,x)} \mathfrak{D}(\mathbb{Y}_{i,j}(N)^x).$$

The RHS represents a sum of distances of columns obtained by removing a set of rows $M$. For each such $M \subset \{1, \ldots k\}$, any $x \in M$ produces a unique $N = M \setminus \{x\}$. Consequently each $M$ is induced by exactly $s$ pairs $(N, x)$ and we obtain

$$\frac{(s-1)!(k+1-s)!}{k!(k+1-s)} \sum_{(N,x)} \mathfrak{D}(\mathbb{Y}_{i,j}(N)^x) = \frac{1}{\binom{k}{s}} \sum_M \mathfrak{D}(\mathbb{Y}_{i,j}(M)) = \mathbb{E}_s\left[\mathfrak{D}(\mathbb{Y}_{i,j})\right].$$

Thus we have shown

$$\mathbb{E}_{s-1}\left[\mathfrak{D}(\mathbb{Y}_{i,j})\right] - \mathbb{E}_s\left[\mathfrak{D}(\mathbb{Y}_{i,j})\right] \geq 0. \tag{3.9}$$

Combining eq. (3.9) with the induction hypothesis in eq. (3.6), we derive

$$\left[\mathfrak{D}(Y_{i,j}) - \mathbb{E}_{s-1}\left[\mathfrak{D}(\mathbb{Y}_{i,j})\right]\right] + \left[\mathbb{E}_{s-1}\left[\mathfrak{D}(\mathbb{Y}_{i,j})\right] - \mathbb{E}_s\left[\mathfrak{D}(\mathbb{Y}_{i,j})\right]\right] > 0. \tag{3.10}$$

In view of eq. (3.10) and linearity of expectation, we therefore have proved

$$\mathbb{E}_s[\mathcal{Y}_{i,j}] = \mathbb{E}_s[\mathfrak{D}(Y_{i,j}) - \mathfrak{D}(\mathbb{Y}_{i,j})] = \mathfrak{D}(Y_{i,j}) - \mathbb{E}_s[\mathfrak{D}(\mathbb{Y}_{i,j})] > 0.$$

The second assertion of the theorem is an immediate consequence of the fact that $\mathbb{E}_s[\mathcal{Y}_{i,j}] > 0$ in case the link is not pure, completing the proof of the theorem. $\square$

## 4. The ground state

A pair of columns $Y_{i,j}$, in which the symbol pair $(y_i, y_j) \in \mathcal{A}^2$ appears with equal frequency, is denoted by $U_{i,j}$ and called uniform. Then, by abuse of notation, we shall also refer to the link $(i, j)$ as uniform. Clearly, the existence of uniform links is equivalent to $k \equiv 0 \mod a^2$ where $a = |\mathcal{A}|$. For such $k$, we have $\mathfrak{D}(Y_{i,j}) \leq 2\ln(a)$, with equality, if and only if $Y_{i,j} = U_{i,j}$ [20], i.e. uniform links exhibit maximal variation of information distance.

In this section we will assume that $k$ satisfies $k \equiv 0 \mod a^2$ and show that the link-potential for links that are not pure, assumes its minimum only for uniform links. This minimum depends on $k$ and strictly decreases as a function of $k$.

**Theorem 4.1.** *Let $Y_{i,j}$ be a pair of columns of size $k \equiv 0 \mod a^2$, for which the link $(i, j)$ is not pure. Then*

$$\mathbb{E}_1[\mathcal{Y}_{i,j}] \geq \frac{2}{k-1}\left[g\left(\frac{k}{a}\right) - g\left(\frac{k}{a^2}\right)\right]. \tag{4.1}$$

*Furthermore, the lower bound of eq. (4.1) is sharp and exclusively assumed in case $Y_{i,j} = U_{i,j}$.*

To prove Theorem 4.1 we first establish a technical lemma:

**Lemma 4.1.** *(a) $G(z)$ is positive, concave for $1 \leq z$, (b) $H_a(z) = G(z) - aG\left(\frac{z}{a}\right)$ is positive and strictly convex for $2 \leq a < z$.*

*Proof.* To prove (a), we show $G''(z) \leq 0$. To this end we compute

$$G'(z) = (2z-1)\ln\left[1 + \frac{1}{z-1}\right] - 1, \quad G''(z) = \frac{1-2z}{z(z-1)} + 2\ln\left[1 + \frac{1}{z-1}\right].$$

Setting $z = (1 + \epsilon)$, where $\epsilon \geq 0$ we derive

$$G''(1 + \epsilon) = -\frac{1 + 2\epsilon}{\epsilon(1 + \epsilon)} + 2\ln\left[1 + \frac{1}{\epsilon}\right].$$

In view of [24], the term $\ln(1 + x)$ is bounded as follows

$$\frac{2x}{2+x} \leq \ln(1+x) \leq \frac{x}{2}\frac{2+x}{1+x}. \tag{4.2}$$

Using the RHS of eq. (4.2) and setting $x = 1/\epsilon$, assertion (a) follows from

$$G''(1+\epsilon) \leq -\frac{1+2\epsilon}{\epsilon(1+\epsilon)} + 2\frac{1/\epsilon}{2}\frac{2+1/\epsilon}{1+1/\epsilon} = -\frac{1+2\epsilon}{\epsilon(1+\epsilon)} + \frac{1+2\epsilon}{\epsilon(1+\epsilon)} = 0.$$

To show (b), we first introduce

$$F_z(a) := \frac{1}{z-a} - \frac{2}{a} \ln \left[ \frac{z}{z-a} \right].$$

Then

$$
\begin{aligned}
H_a''(z) &= G''(z) - \frac{1}{a} G'' \left( \frac{z}{a} \right) \\
&= \frac{1-2z}{z(z-1)} + 2\ln \left[ 1 + \frac{1}{z-1} \right] - \frac{1}{a} \left[ \frac{1-2z/a}{z/a(z/a-1)} + 2\ln \left[ 1 + \frac{1}{z/a-1} \right] \right] \\
&= -\frac{1}{z} - \frac{1}{z-1} + 2\ln \left[ 1 + \frac{1}{z-1} \right] - \frac{1}{a} \left[ \frac{1}{z/a} - \frac{1}{z/a-1} \right] - \frac{2}{a} \ln \left[ 1 + \frac{1}{z/a-1} \right] \\
&= \underbrace{-\frac{1}{z-1} + 2\ln \left[ 1 + \frac{1}{z-1} \right]}_{-F_z(1)} + \underbrace{\frac{1}{z-a} - \frac{2}{a} \ln \left[ 1 + \frac{1}{z/a-1} \right]}_{F_z(a)}.
\end{aligned}
$$

Accordingly $H_a''(z) = F_z(a) - F_z(1)$, where $F_z(a)$ is differentiable as a function of $a$, for any values $a < z$. The mean value theorem then implies

$$
\begin{aligned}
H_a''(z) &= F_z'(\xi)(a-1) \\
&= \left[ (-1)\frac{-1}{(z-\xi)^2} + \frac{2}{\xi^2} \ln \left[ \frac{z}{z-\xi} \right] - \frac{2}{\xi} \frac{z-\xi}{z} (-1) \frac{z}{(z-\xi)^2} (-1) \right] (a-1),
\end{aligned}
$$

where $1 < \xi < a$. This is equivalent to

$$H_a''(z) = \frac{\xi^2 - 2\xi(z-\xi) + 2(z-\xi)^2 \ln \left[ 1 + \frac{\xi}{z-\xi} \right]}{\xi^2 (z-\xi)^2} (a-1).$$

The expressions $\xi^2(z-\xi)^2$, $\ln \left[ 1 + \frac{\xi}{z-\xi} \right]$ as well as $(a-1)$ are positive and employing eq. (4.2) we obtain the lower bound for the numerator

$$\xi^2 - 2\xi(z-\xi) + 2(z-\xi)^2 \frac{2\frac{\xi}{z-\xi}}{2 + \frac{\xi}{z-\xi}} \le \xi^2 - 2\xi(z-\xi) + 2(z-\xi)^2 \ln \left[ 1 + \frac{\xi}{z-\xi} \right]. \tag{4.3}$$

The LHS of eq. (4.3) equals $\xi^2 - 2\xi(z-\xi) + 4(z-\xi)^2 \frac{\xi}{2z-\xi}$ and in view of $\xi^2(2z-\xi) - 2\xi(z-\xi)(2z-\xi) + 4(z-\xi)^2\xi = \xi^3$, we arrive at

$$0 < \frac{\xi^3(a-1)}{\xi^2(z-\xi)^2(2z-\xi)} \le H_a''(z),$$

whence (b) and the lemma follows. $\qquad \square$

We are now in position to prove Theorem 4.1.

*Proof.* We shall first compute the link-potential of a uniform system, $\mathbb{E}_1[\mathcal{U}_{i,j}] = \mathbb{E}_1[\mathfrak{D}(U_{i,j}) - \mathfrak{D}(\mathbb{U}_{i,j})]$. Secondly we show that the link-potential of uniform links assumes the unique minimum.

*Claim* 1.

$$\mathbb{E}_1[\mathcal{U}_{i,j}] = \frac{2}{k-1} \left[ g \left( \frac{k}{a} \right) - g \left( \frac{k}{a^2} \right) \right].$$

To prove Claim 1, we observe that in Lemma 3.1, eq. (3.2) implies

$$\mathbb{E}_1[\mathcal{U}_{i,j}] = \frac{\lambda(\mathcal{U}_{i,j})}{k-1} = \frac{1}{(k-1)} \frac{1}{k} \left[ \sum_{y_i} G(\nu_{y_i}) + \sum_{y_j} G(\nu_{y_j}) - 2 \sum_{(y_i, y_j)} G(\nu_{y_i, y_j}) \right].$$

Since for $U_{i,j}$, $\nu_{y_i} = \nu_{y_j} = \frac{k}{a}$, $\nu_{y_i,y_j} = \frac{k}{a^2}$ holds, Claim 1 immediately follows from

$$\frac{1}{k} \sum_{y_h} G(\nu_{y_h}) = \frac{1}{k} \cdot a \cdot \frac{k}{a} \cdot g\left(\frac{k}{a}\right), \quad \frac{1}{k} \sum_{(y_i,y_j)} G(\nu_{y_i,y_j}) = \frac{1}{k} \cdot a^2 \cdot \frac{k}{a^2} \cdot g\left(\frac{k}{a^2}\right).$$

where $h \in \{i, j\}$.

*Claim* 2. For any $Y_{i,j} \neq U_{i,j}$ we have

$$\mathbb{E}_1[\mathcal{Y}_{i,j}] > \mathbb{E}_1[\mathcal{U}_{i,j}].$$

Let

$$\mathbb{E}_1[\mathcal{Y}_{i,j}] - \mathbb{E}_1[\mathcal{U}_{i,j}] = \frac{1}{k-1}\left[\lambda(Y_{i,j}) - \lambda(U_{i,j})\right] = \frac{\Delta}{k-1},$$

and setting $\gamma = \frac{k}{a^2}$ we consider

$$k\Delta = \sum_{y_i}[G(\nu_{y_i}) - G(a\gamma)] + \sum_{y_j}[G(\nu_{y_j}) - G(a\gamma)] - 2\sum_{(y_i,y_j)}[G(\nu_{y_i,y_j}) - G(\gamma)].$$

The symmetry in $y_i, y_j$ reduces the analysis to establishing

$$T_{y_h} = \left[\sum_{y_h} G(\nu_{y_h}) - aG(a\gamma) - \sum_{(y_i,y_j)} G(\nu_{y_i,y_j}) + a^2 G(\gamma)\right] > 0, \quad h \in \{i, j\}$$

and it suffices to consider $h = i$. Since $a > 0$, strict positivity of $\Delta$ is guaranteed by

$$\frac{1}{a}T_{y_i} = \left[\sum_{y_i} \frac{1}{a}G(\nu_{y_i}) - \sum_{y_i}[\sum_{y_j} \frac{1}{a}G(\nu_{y_i,y_j})] + aG(\gamma) - G(a\gamma)\right] > 0.$$

For $z \geq 1$, Lemma 4.1 establishes that $G(z)$ is positive and concave. We apply Jensen's inequality and derive

$$\sum_{y_j} \frac{1}{a}G(\nu_{y_i,y_j}) = \frac{\sum_{y_j} \frac{1}{a}G(\nu_{y_i,y_j})}{\sum_{y_j} \frac{1}{a}} \leq G\left(\frac{\sum_{y_j} \frac{1}{a}\nu_{y_i,y_j}}{1}\right) = G\left(\frac{1}{a}\nu_{y_i}\right).$$

Consequently, Jensen's inequality eliminates the summation over $y_j$, and it suffices to show

$$\left[\frac{1}{a}\sum_{y_i}\left[G(\nu_{y_i}) - aG\left(\frac{1}{a}\nu_{y_i}\right)\right]\right] - G(a\gamma) + aG(\gamma) > 0. \tag{4.4}$$

In view of $H_a(z) = G(z) - aG(\frac{z}{a})$ we can interpret eq. (4.4) as follows

$$\left[\frac{1}{a}\sum_{y_i}\left[G(\nu_{y_i}) - aG\left(\frac{1}{a}\nu_{y_i}\right)\right]\right] - G(a\gamma) + aG(\gamma) = \left[\frac{1}{a}\sum_{y_i} H_a(\nu_{y_i})\right] - H_a(a\gamma). \tag{4.5}$$

In view of Lemma 4.1 (b), $H_a(z)$ is positive and strictly convex. Since the distribution $(p_{y_i})_{y_i \in \mathcal{A}}$ is non-degenerate, Jensen's inequality holds as a strict inequality

$$H_a\left(\frac{k}{a}\right) = H_a\left(\frac{\sum_{y_i} \frac{1}{a}\nu_{y_i}}{\sum_{y_i} \frac{1}{a}}\right) < \frac{\sum_{y_i} \frac{1}{a}H_a(\nu_{y_i})}{\sum_{y_i} \frac{1}{a}} = \sum_{y_i} \frac{1}{a}H_a(\nu_{y_i}).$$

Expressing the uniform marginals as $\frac{k}{a} = a\gamma$ and substituting the bound in eq. (4.5) we arrive at

$$0 = H_a\left(\frac{k}{a}\right) - H_a(a\gamma) < \left[\frac{1}{a}\sum_{y_i} H_a(\nu_{y_i})\right] - H_a(a\gamma),$$

whence then claim, and the theorem is proved. □

# 5. The site linkage spectrum

Having established basic properties of the link-potential, $\mathbb{E}_s[\mathcal{Y}_{i,j}]$, we are now in position to develop first aspects of an information theoretic framework for the analysis of aligned data.

Given a system, $\Sigma = Y_{1,\dots,n}$, of size $k$ and length $n$ with entries in $\mathcal{A}$, we denote its associated set of links by $E := \{(i,j) \mid 1 \leq i < j \leq n\}$. Then, we define the *site linkage spectrum* of $\Sigma$ as the multi-set of potentials of the $E'$-links, where $E' \subseteq E$,

$$\mathfrak{S}_s(\Sigma, E') := \{\mathbb{E}_s[\mathcal{Y}_{i,j}] \mid (i,j) \in E'\}.$$

In the following we fix a system $\Sigma$ together with $E' \subset E$ and employ the site-linkage spectrum in order to derive a partition of $\Sigma$ into blocks of sequences. These blocks are comprised of sequences that exhibit a certain information theoretic coherence. We shall show that this line of work can be used to identify "outlier" sequences within a given alignment.

The key quantity here is the effect of removing a single row from $\Sigma$ on the average $\mathfrak{D}$-distance among the column pairs represented by $E'$. Theorem 4.1 shows that this effect is minimized at uniform links, even though uniform links feature a maximal $\mathfrak{D}$-distance. Furthermore, Theorem 4.1 establishes that all row removals have equal effect. At uniform links the average variation in information is minimal and we have isotropy w.r.t. row removal.

Let

$$q(E', \Sigma) = \frac{1}{|E'|} \sum_{(i,j) \in E'} \mathfrak{D}(Y_{i,j}),$$

where $q(E', \Sigma)$ is non negative, and it is zero if and only if all $E'$-links are pure.

We next introduce

$$Q(x) = q(E', \Sigma) - q(E', \Sigma^x), \quad Q(x^*) = \max_x \{Q(x)\},$$

which allows us to identify the set of distinguished rows $x^*$ that maximize the drop in average variation of information. In light of Theorem 3.1, removing the row $x^*$ has the effect of producing a sub-alignment that has, in some sense, become more pure.

While there does not exist a row that simultaneously lowers each term in $q(E', \Sigma) \geq 0$, Theorem 3.1 will guarantee that, in the presence of non-pure $E'$-links, there exists some row, $x^*$, that strictly lowers $q(E', \Sigma)$ as a whole, i.e. $q(E', \Sigma^{x^*}) < q(E', \Sigma)$. Furthermore, if $k \equiv 0 \mod a^2$, Theorem 4.1 will guarantee a minimum increase in purity proportional to the fraction of non-pure links in $E'$:

**Theorem 5.1.** *Fix $E' \subset E$, let $p$ denote the fraction of pure links of $E'$, and suppose $p < 1$. Then there exists some row, $x^*$, such that*

$$Q(x^*) > 0.$$

*Furthermore, if $k \equiv 0 \mod a^2$, then*

$$Q(x^*) \geq \frac{2(1-p)}{k-1} \left[ g\left(\frac{k}{a}\right) - g\left(\frac{k}{a^2}\right) \right] > 0.$$

*Proof.*

$$Q(x^*) \geq \frac{1}{k} \sum_x [q(E', \Sigma) - q(E', \Sigma^x)] = \frac{1}{k|E'|} \left[ \sum_x \sum_{(i,j) \in E'} \mathfrak{D}(Y_{i,j}) - \sum_x \sum_{(i,j) \in E'} \mathfrak{D}(Y_{i,j}^x) \right]. \tag{5.1}$$

Changing the order of summation in the RHS of eq. (5.1), we observe that the latter represents the average of $\mathfrak{S}_1(\Sigma, E')$. By Theorem 3.1 we have

$$\frac{1}{k|E'|} \left[ \sum_x \sum_{(i,j) \in E'} \mathfrak{D}(Y_{i,j}) - \sum_x \sum_{(i,j) \in E'} \mathfrak{D}(Y_{i,j}^x) \right] = \frac{1}{|E'|} \sum_{(i,j) \in E'} \mathbb{E}_1[\mathcal{Y}_{i,j}] > 0.$$

where the last inequality holds because $p < 1$. For the second assertion, by Theorem 4.1

$$\frac{1}{|E'|} \sum_{(i,j) \in E'} \mathbb{E}_1[\mathcal{Y}_{i,j}] \geq (1-p) \mathbb{E}_1[\mathcal{U}_{i,j}] = \frac{2(1-p)}{k-1} \left[ g\left(\frac{k}{a}\right) - g\left(\frac{k}{a^2}\right) \right] > 0,$$

and the theorem follows. $\qquad\square$

Theorem 5.1 guarantees that starting with $\Sigma$, we can iteratively remove rows that produce $E'$-links that become more pure–until, say, a certain threshold is reached, see Figure 1. We shall next illustrate the application of site linkage spectra in a particular case, where we consider single row removals. Let $\Sigma$, be of size $k = 100$ and length $n = 99$, constructed from inverse-fold solutions of two distinct tRNA structures. $\Sigma$ was constructed as follows: firstly, the consensus secondary structure of the tRNA molecules across all species in the tRNAdb [25] was chosen. This secondary structure contains 4 helixes, 21 base pairs, and 57 unpaired bases. A random secondary structure was then sampled having the same length, number of base pairs and number of helixes. The sequence alignment is comprised of 85 sequences obtained by inverse folding the tRNA structure and 15 sequences obtained by inverse folding the alternative structure. The inverse folding algorithm ViennaRNA [26,27] was employed without considering $GU$ base pairs. For each sequence, $x$, the gain in purity was computed when passing to $\Sigma^x$, w.r.t. the fixed link set $E' \subset E$. This link set was ranked by linearly sorting $E$ in ascending order, according to their $\mathfrak{D}$-distances in $\Sigma$, and then including in $E'$ the links belonging to the first $0.5\%$ ranks.

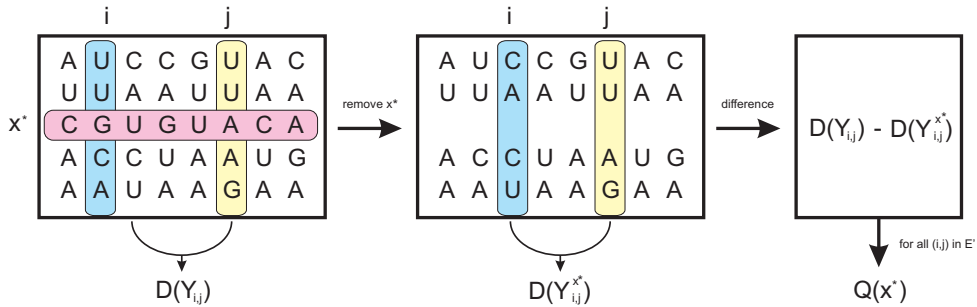

**Figure 1.** Computing $Q(x^*)$ for the particular row index $x^*$ for fixed set of links $E'$.

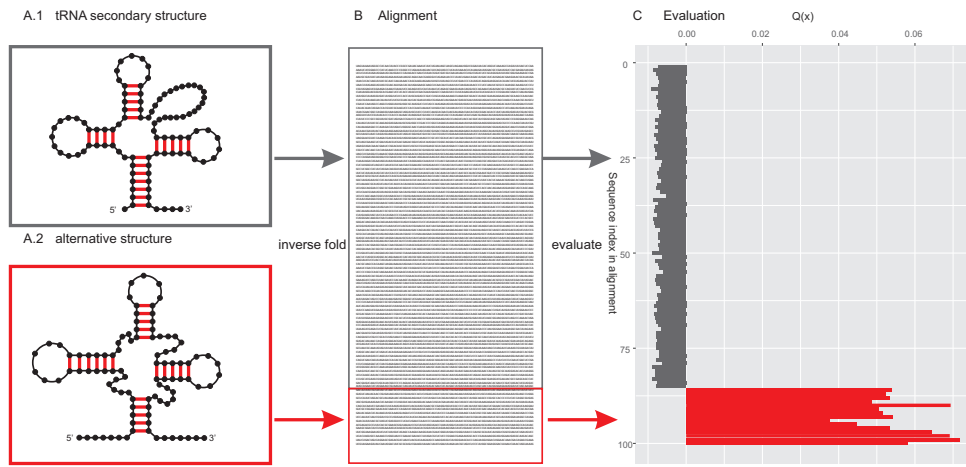

**Figure 2.** (A.1) The consensus secondary structure of tRNA molecules across species in tRNAdb. (A.2) An alternative secondary structure having the same length, number of base pairs and number of helixes as (A.1). (B) A sequence alignment composed of 85 sequences (grey box) obtained by inverse folding (A.1), and 15 sequences (red box) obtained by inverse folding(A.2) (C) Evaluation of $Q(x)$, where $x$ is the respective sequence index in the alignment (B), with respect to a link set $E'$ consisting of the lowest $0.5\%$ $\mathcal{D}$-distance links.

**Acknowledgements.** The authors would like to acknowledge discussions with Michael Waterman that have improved the readability of the manuscript.

Funding provided in part under VACA, #HHM402-23-C-0028.

The authors of this paper declare no competing interests.